# Charge-Stripe Crystal Phase in an Insulating Cuprate

He Zhao[1], Zheng Ren[1], Bryan Rachmilowitz[1], John Schneeloch [2], Ruidan Zhong[2], Genda Gu[2], Ziqiang Wang[1] and Ilija Zeljkovic[1,*]

[1]Department of Physics, Boston College, 140 Commonwealth Ave, Chestnut Hill, MA 02467

[2]Brookhaven National Laboratory, Upton, NY 11973

[*]Correspondence to: ilija.zeljkovic@bc.edu

**High-$T_c$ superconductivity in cuprates is generally believed to arise from carrier doping an antiferromagnetic Mott (AFM) insulator. Theoretical proposals and emerging experimental evidence suggest that this process leads to the formation of intriguing electronic liquid crystal phases [1]. These phases are characterized by ordered charge and/or spin density modulations, and thought to be intimately tied to the subsequent emergence of superconductivity. The most elusive, insulating charge-stripe crystal phase is predicted to occur when a small density of holes is doped into the charge-transfer insulator state [1–3], and would provide a missing link between the undoped parent AFM phase and the mysterious, metallic "pseudogap". However, due to experimental challenges, it has been difficult to observe this phase. Here, we use surface annealing to extend the accessible doping range in Bi-based cuprates and achieve the lightly-doped charge-transfer insulating state of a cuprate $Bi_2Sr_2CaCu_2O_{8+x}$. In this insulating state with a charge transfer gap at the order of ~1 eV, using spectroscopic-imaging scanning tunneling microscopy, we discover a unidirectional charge-stripe order with a commensurate $4a_0$ period along the Cu-O-Cu bond. Importantly, this insulating charge-stripe crystal phase develops before the onset of the pseudogap and the formation of the Fermi surface. Our work provides new insights into the microscopic origin of electronic inhomogeneity in high-$T_c$ cuprates.**

Uncovering the quantum electronic states connecting the two known phases of the cuprate superconductors – the undoped antiferromagnetic Mott (AFM) insulator with a charge transfer gap and the *d*-wave superconductor at sufficiently high carrier doping – has been the central challenge for understanding unconventional superconductivity [4]. The experiments so far have primarily focused on exploring novel forms of electronic order in the pseudogap phase of hole-doped cuprates, before and shortly after the onset of superconductivity. Pioneering scattering studies probing the charge dynamics in La- and Y-based cuprates discovered static [5–7] as well as fluctuating [8] charge order phase that is also unidirectional ("stripe") [5–9], in analogy to the electronic liquid crystal nematic or smectic phases proposed for frustrated phase separation [1]. Experiments on Bi- and Cl-based cuprates also revealed a charge order in the pseudogap phase [10–23], but whether the order originated in the bi-directional "checkerboard" [16,17] or the disorder-pinned stripes [10,18,20,21] remains debated. Moreover, the link between these charge-ordered states and the parent charge-transfer insulator state before the onset of the pseudogap remains poorly understood. A key initial proposal suggested that doping induces an inhomogeneous ground state with a tendency to form periodic charge and spin stripes even in the lightly-doped AFM phase [2], an ordering possibly stabilized by the long range Coulomb interaction [3]. However, a direct evidence for such insulating stripe crystal phase in the insulating cuprates when the charge transfer gap is on the order of ~1 eV has been challenging to observe thus far, in large part due to the difficulty in synthesizing and characterizing bulk cuprates in the very lightly doped insulating regime at low temperatures [13,24].

$Bi_2Sr_2CaCu_2O_{8+x}$ (Bi-2212) is a layered cuprate in which the hole density is controlled by off-stoichiometric (interstitial) oxygen dopant atoms (Fig. 1A), each expected to contribute ~2 holes to the bulk. As-grown Bi-2212 single crystals tend to be optimally-doped (hole density $p$ ~ 0.16 per unit cell) with superconducting transition temperature of $T_c$ ~ 91 K. To lower $p$, these samples can be subsequently annealed at high temperature in ultra-high vacuum (UHV), which removes a fraction of the interstitial oxygen atoms from the bulk. If this process is performed over the course of days, it can ultimately yield bulk insulating samples close to the AFM transition. However, these have proven to be difficult to characterize via tunneling techniques due to insufficient charge carriers available for tunneling [13,24]. Here, we circumvent this problem by demonstrating that a short cycle of annealing in UHV can lead to a significant decrease in the hole density primarily near the exposed topmost surface, while the bulk remains conducting (Supplementary Information 3), providing enough charge carriers for tunneling experiments. This breakthrough allows us to use scanning tunneling microscopy and spectroscopy (STM/S) to study the previously inaccessible insulating state of Bi-2212 at ~4.5 K, in the regime before the charge transfer gap is closed. We find that when only a small density of holes is doped into Bi-2212, an insulating charge-ordered phase arises with commensurate $4a_0$ unidirectional charge density modulations along the Cu-O-Cu bond (where $a_0$ ~ 3.8 Å corresponds to the Bi-Bi, or equivalently Cu-Cu separation).

We start our experiment by characterizing the surface of UHV-cleaved optimally-doped Bi-2212 bulk single crystals ($T_c$ ~ 91 K), prior to any annealing. Consistent with previous reports and the approximate doping level [25,26], $dI/dV$ spectra (where $I$ is the current and $V$ is the voltage applied to the sample) exhibit a spatially inhomogeneous spectral gap, with the average magnitude $\Delta_0$ ~ 43 meV calculated as one half of the energy difference between the two gap edge peaks (Fig. 1B). The same sample is then annealed at ~270 °C in UHV (Methods) and re-inserted into the STM. Although the STM topograph of the post-annealed surface looks qualitatively similar (inset in Fig. 1C), the average $dI/dV$ spectrum is strikingly different (Fig. 1C). The gap-edge peaks at $\Delta_0$ are now completely suppressed, while the emergent V-shaped spectrum with poorly defined gap-edge peaks at higher energies (Fig. S6H) is reminiscent of that obtained on underdoped Bi-based cuprates with low, but finite $T_c$ [15]. Interestingly, after annealing the sample at even higher temperature up to ~380 °C, we detect a large insulating gap $\Delta_i$ ~ 1.1 eV in $dI/dV$ spectra (Fig. 1D), indicating a transition into the insulating phase (Fig. 1F).

In order to quantify this process, we first use high-bias imaging, which allows us to visualize oxygen dopants responsible for hole doping as bright features in STM $dI/dV(\mathbf{r},V)$ maps (where $\mathbf{r}$ is the lateral position on the sample) [27,28]. Bi-2212 harbors two main types of interstitial O atoms – "type-A" O located near the SrO plane and "type-B" O positioned closer to the BiO surface layer [27–29]. In the post-annealed sample in Fig. 1C, we find that the densities of both types of O interstitials are significantly lower than the concentrations in the optimally-doped sample prior to annealing (Fig. 1E, Supplementary Information 2). This confirms the expected scenario of oxygen interstitials being removed from the sample by heating. The annealing process also creates a small number of apical oxygen vacancies in the SrO layer (Fig. 1E), which are commonly found in underdoped bulk Bi-2212 [28] and expected to further underdope the sample. By comparing the concentration of different types of O dopants measured here with those reported in Ref. [28], we estimate the hole density of the sample annealed at ~270 °C in Fig. 1C to be $p$ ~ 0.057±0.008

(Supplementary Information 2). This is consistent with the average V-shaped *dI/dV* spectrum (Fig. 1C) being qualitatively similar to that obtained on underdoped bulk Bi-2212 with $p \sim 0.063$ [15]. Moreover, *dI/dV*(**r**,V) maps at the bias near the Fermi level display the familiar charge order (Fig. S6) [12,22,23], with the incommensurate wave vectors **Q**\* = $0.28 \pm 0.03$ and **Q**\*\* = $0.73 \pm 0.04$ (we hereafter define reciprocal lattice vector $2\pi/a_0 = 1$) along both lattice directions, consistent with those measured on the samples with similar bulk hole density [23].

Previous observations of charge ordering in hole-doped cuprates [10–23] have been limited to the approximate doping level achieved in Fig. 1C, in the pseudogap regime with a finite density of states at the Fermi level, after the charge transfer gap had been closed. Now, we turn to the characterization of the insulating Bi-2212 surface, which shows a large insulating gap $\Delta_i \sim 1.1$ eV in the average *dI/dV* spectrum at ~4.5 K (Fig. 1D, Fig. S10), comparable to that measured on bulk insulating Bi-2212 at ~77 K [24]. The gap is asymmetric with respect to the Fermi level and extends from approximately -0.3 eV to 0.8 eV. Next, we acquire high-resolution STM topographs in the constant-current mode, which inevitably contain both the electronic and structural information. A typical STM topograph of the insulating sample clearly shows the individual Bi atoms on the surface as well as the characteristic supermodulation oriented at a 45° angle with respect to the lattice (Fig. 2A). Remarkably, the topograph also exhibits periodic unidirectional features aligned parallel to the Cu-O-Cu lattice direction (Fig. 2A), which have not been observed before. We rule out the structural origin of these modulations by examining reflection high-energy electron diffraction patterns of the post-annealed surface that look identical to the one before heating (Supplementary Information 1). The spatial extent of unidirectional stripe domains is typically ~5-10 nm in length, but it can be as large as a ~15 nm square region in Fig. 2C, spanning >1,000 unit cells. Their periodic nature can be confirmed by examining discrete two-dimensional Fourier transform (FT) of the entire STM topograph, where two peaks labeled $\mathbf{Q}^X_{CO}$ and $\mathbf{Q}^Y_{CO}$ emerge at (0.25, 0) and (0.25, 0), respectively (Fig. 2B). These wave vectors suggest that the real-space period of the modulations is exactly $4a_0$ along either lattice directions. Based on the filtered topograph in Fig. 2E and associated FT linecuts that show in-phase relationship between the atomic Bragg peak $\mathbf{Q}_x$ and $\mathbf{Q}^X_{CO}$ (Fig. 2F), we further conclude that the peaks of the stripe modulation are site-centered, positioned on top of Bi and therefore the Cu sites. This is in contrast to the bond-centered charge modulations observed in underdoped cuprates in the pseudogap regime [18].

The unidirectional nature of the modulation can be clearly visualized as long horizontal lines that run across the entire field-of-view (Fig. 2C), overcoming the structural superlattice modulations ($Q_{SM}$ in Fig. 2B). Quantitatively, unidirectionality within a single stripe domain can also be confirmed by the presence of only $Q^X_{CO}$ (but not $Q^Y_{CO}$) in the FT of the STM topograph (Fig. 2D, 2F). Moreover, by constructing the modulation amplitude maps along **x** (Fig. 2G) and **y** lattice directions (Fig. 2H) over a large area, we can observe a strong anti-correlation between the two (inset in Fig. 2H), a trend also confirmed for $dI/dV(\mathbf{r},V)$ maps (Fig. S8). This is consistent with the visual observation that a certain region of the sample hosts either the modulation along **x** or along **y** lattice direction, but not both, implying a unidirectional stripe order. The unidirectional nature cannot be explained by STM tip anisotropy because we observe stripes oriented along both lattice directions in the same field-of-view using the same tip (Fig. 2A).

We further investigate the electronic origin of this modulation by acquiring $dI/dV(\mathbf{r},V)$ maps, where the modulation is visible as bright lines directly corresponding to the equivalent features observed in the STM topograph (Fig. 3A,B). The associated FT also shows peaks at $Q^X_{CO}$ ~(0.25,0) and $Q^Y_{CO}$ ~(0,0.25) (Fig. 3C), thus demonstrating the ordered nature of the modulations. These modulations are non-dispersive in a wide range of STM biases (Fig. 3D, 3E), which is consistent with charge ordering. The same ordering vector is also seen in normalized $R(\mathbf{r},|V|)$ maps used to remove the STM setup condition [19]. To describe the electronic structure of the striped hole crystal state in more detail, the periodic variation in $dI/dV(\mathbf{r}, V)$ conductance is shown in Fig. 3F along a linecut across the stripes indicated in Fig. 3D. This modulation in the ordered regions by the $4a_0$ period can also be seen in the spatial map of the charge transfer gap and its associated FT (Figs. S10, S11), which is one of the indications suggesting the importance of $CuO_2$-derived Hubbard bands on the physics of the charge-stripe phase (Supplementary Information 5). Our experiments have not been able to pinpoint what type of chemical disorder, if any, could be related to the observed charge modulations. Annealing at temperatures used in this work is only expected to affect the oxygen dopant density, not the cation concentration [30]. The analysis of oxygen dopant distribution in our surface-annealed samples reveals no spatial ordering with the $4a_0$ period (Supplementary Information 4). In analogy to the observations in the pseudogap phase of Bi-2212 [28], some correlation between oxygen defects and maxima/minima of the charge modulations may be possible.

In addition to previously observed smectic [5–10,13,14,18,20,21,23] and nematic [15] orders in the pseudogap phase, we have reported here the observation of perhaps the most elusive electron liquid crystal phase in lightly hole-doped cuprates – the charge-stripe crystal phase. In sharp contrast to the charge ordered phases in the pseudogap phase manifested at low energies near the Fermi level, the striped hole crystal phase is detected in the insulating state and likely tied to the spatial modulations of the high-energy charge transfer gap. Given the long-standing debate on whether the stripes (checkerboard) that break (preserve) the $C_4$ rotational symmetry of the crystal are favored in cuprates [31,32], our work provides a real-space evidence that clearly indicates the preference of lightly-doped cuprates to form charge stripes over the checkerboard. Our observation of large phase-coherent regions indicates that the striped hole crystal may exist as a stable ground state of a lightly-doped Mott insulator stabilized by long-range Coulomb interaction [3]. Moreover, the coexistence of ordered and disordered regions within the same field-of view (for example solid and dashed squares in Fig. 3D) could be indicative of Coulomb frustrated phase separation. The observed charge-stripe phase is also consistent with the picture of fluctuating stripes [1,33,34] pinned by chemical disorder. We expect it could be detected by resonant X-ray scattering [12,23] provided that the insulating state achieved at the surface by annealing extends into the bulk over micrometer-scale distances. We note that our STM data do not resolve the role of the spin in the observed stripe hole crystal. However, since insulating stripes with both charge and spin order have been observed in the hole-doped nickelate $La_2NiO_{4+x}$ [35], albeit being diagonal and with a much larger charge transfer gap of ~4 eV, it is possible that the insulating charge stripes in Bi-2212 also carry a periodic ordering of spins. Our work provides a viable platform to test this using spin-polarized STM.

Recent experiments have suggested that the charge ordering in cuprates may be intimately related to the pseudogap, and that its wave vector exhibits a doping-dependence as if it was driven by the Fermi surface effects [36]. However, our data challenges this momentum-space picture, as we observe commensurate charge ordering in the charge transfer insulator state, before the onset of the pseudogap and the emergence of a Fermi surface. It is conceivable that the stripe crystal phase serves as the reference insulating state. In this scenario, further hole doping closes the charge transfer gap by nucleating inherently nematic low energy quasiparticle states, which are unstable to the formation of incommensurate charge and other intertwined orders associated with the pseudogap. This is consistent with a more elaborate analysis of STM data, which revealed the underlying commensurate nature of the charge order in the pseudogap phase, disentangled from the apparent incommensurate, doping-dependent vector [14]. However, more experiments are necessary to establish the connections (if any) between the insulating $4a_0$ stripe hole crystal reported here and the charge ordering phases in the pseudogap state. Our work provides a promising direction for exploring this by employing surface annealing.

**Figures**

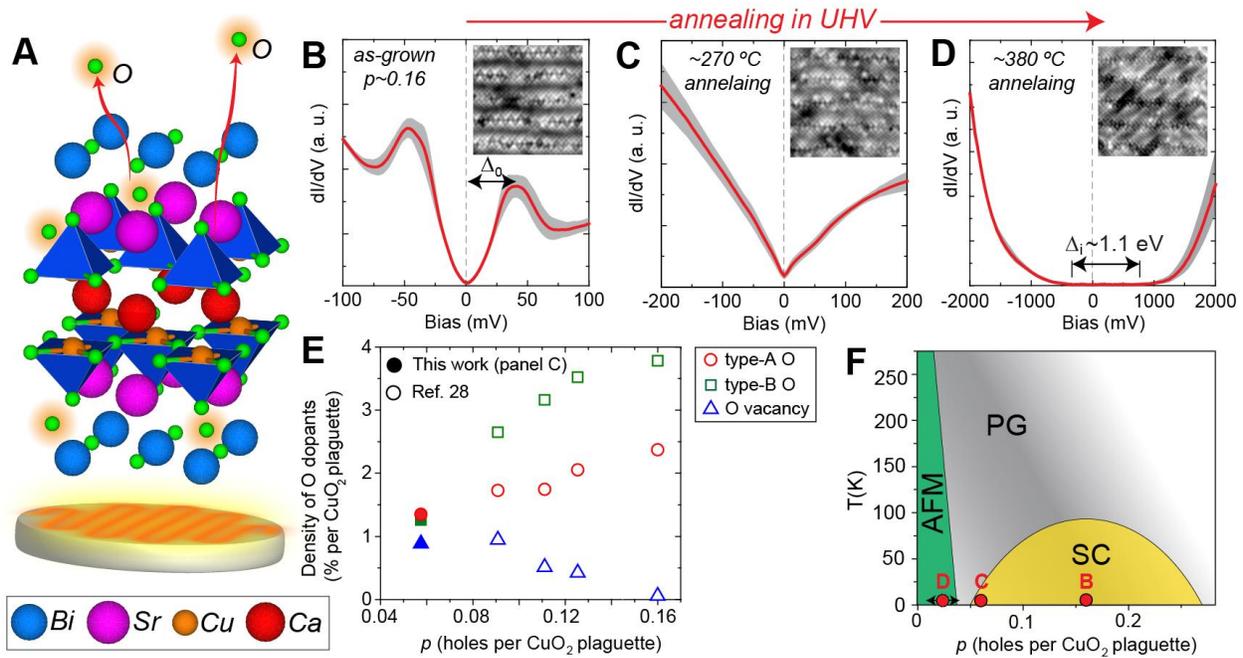

**Figure 1. Surface preparation and electronic characterization**. (A) The schematic of the top half of the Bi-2212 unit cell. Bi-2212 cleaves between two BiO planes to expose a BiO surface measured by STM. Annealing removes a fraction of the oxygen atoms thus reducing the hole density. Average *dI/dV* spectra acquired on the surface (B) before annealing, (C) after annealing at ~270 °C and (D) after annealing at ~380 °C. One half of the width of the gray shaded line at each bias V represents two standard deviations within *dI/dV*(**r**,V). Insets in (B-D) show typical STM topographs. (E) Density of different types of O defects as a function of *p*. (F) Schematic of the hole-doped cuprate phase diagram, with the prominent phases labeled as AFM (antiferromagnetic Mott insulator), SC (superconducting state) and the PG (pseudogap). Red circles represent the approximate doping levels estimated for (B, C) (Figs. S2, S3). The hole density *p* in (D) could not be estimated based on the oxygen defect counting (defects could not be unambiguously identified in *dI/dV*(**r**,V) maps). However, based on the insulating spectra observed, we can schematically represent this data point to be near the AFM transition. STM setup in: (B) $V_{sample}$=-0.1V, $I_{set}$=40pA, $V_{exc}$=4mV (zero-to-peak); (C) $V_{sample}$=0.2V, $I_{set}$=80pA, $V_{exc}$=5mV; (D) $V_{sample}$=-2V, $I_{set}$=20pA, $V_{exc}$=20mV.

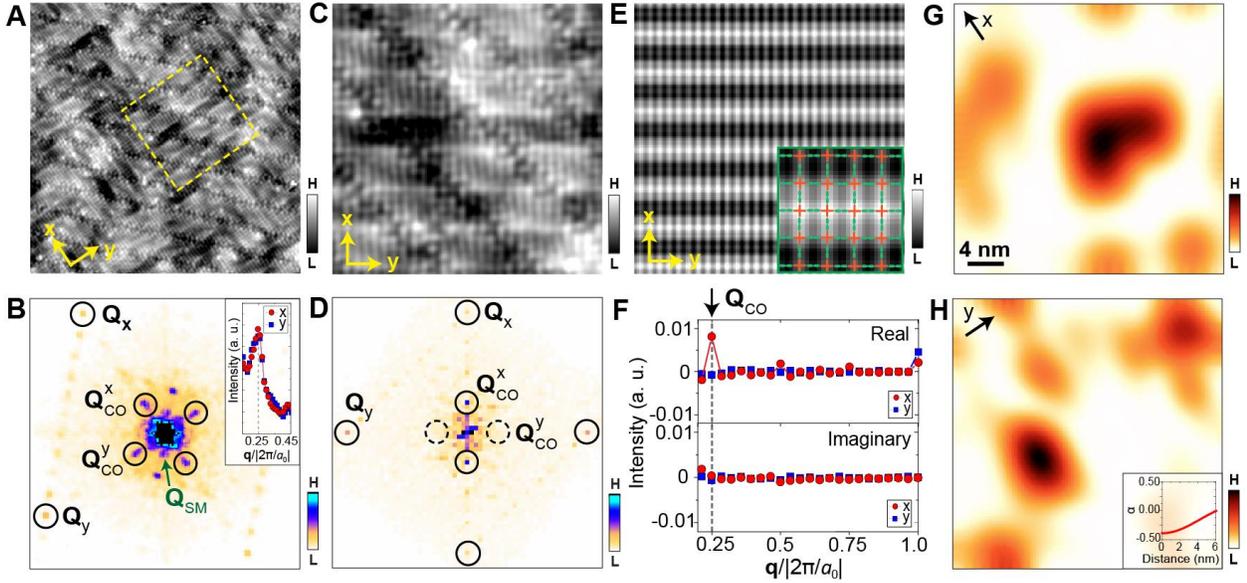

**Figure 2. Unidirectional charge stripe order in insulating Bi-2212.** (A) Atomically-resolved STM topograph of a ~30 nm square region and (B) its associated FT. The atomic Bragg peaks ($Q_x$ and $Q_y$), charge stripe ordering peaks ($Q^X_{CO}$ and $Q^Y_{CO}$) and the structural supermodulation peak ($Q_{SM}$) are denoted. Inset in (B) shows the linecut along $Q_x$ and $Q_y$. (C) Zoom on a ~17 nm square STM topograph hosting a single charge ordering direction (dashed yellow square in (A)), and (D) its associated FT. (E) Fourier-filtered STM topograph from (C) isolating the contributions from wavevectors at $Q_x$, $Q_y$, $Q^X_{CO}$ and $Q^Y_{CO}$, only. The topograph demonstrates that the charge order is site-centered. Inset in (E) shows the position of the Cu atoms (orange crosses) and O atoms (green ellipses) in the $CuO_2$ layer with respect to the filtered STM topograph. (F) Linecuts from the center of the FT to $Q_x$ and $Q_y$, showing the real and the imaginary FT components separately. Both $Q^X_{CO}$ and $Q_x$ show non-zero real components, while the imaginary components are zero, thus quantitatively confirming their in-phase relationship visually observed in (E). The amplitude of the charge stripe order along (G) **x** and (H) **y** lattice directions acquired over the region in (A) that show strong anti-correlation with α ~ -0.4 (inset in H). STM setup in (A, C) $V_{sample}$=-2V and $I_{set}$=3pA.

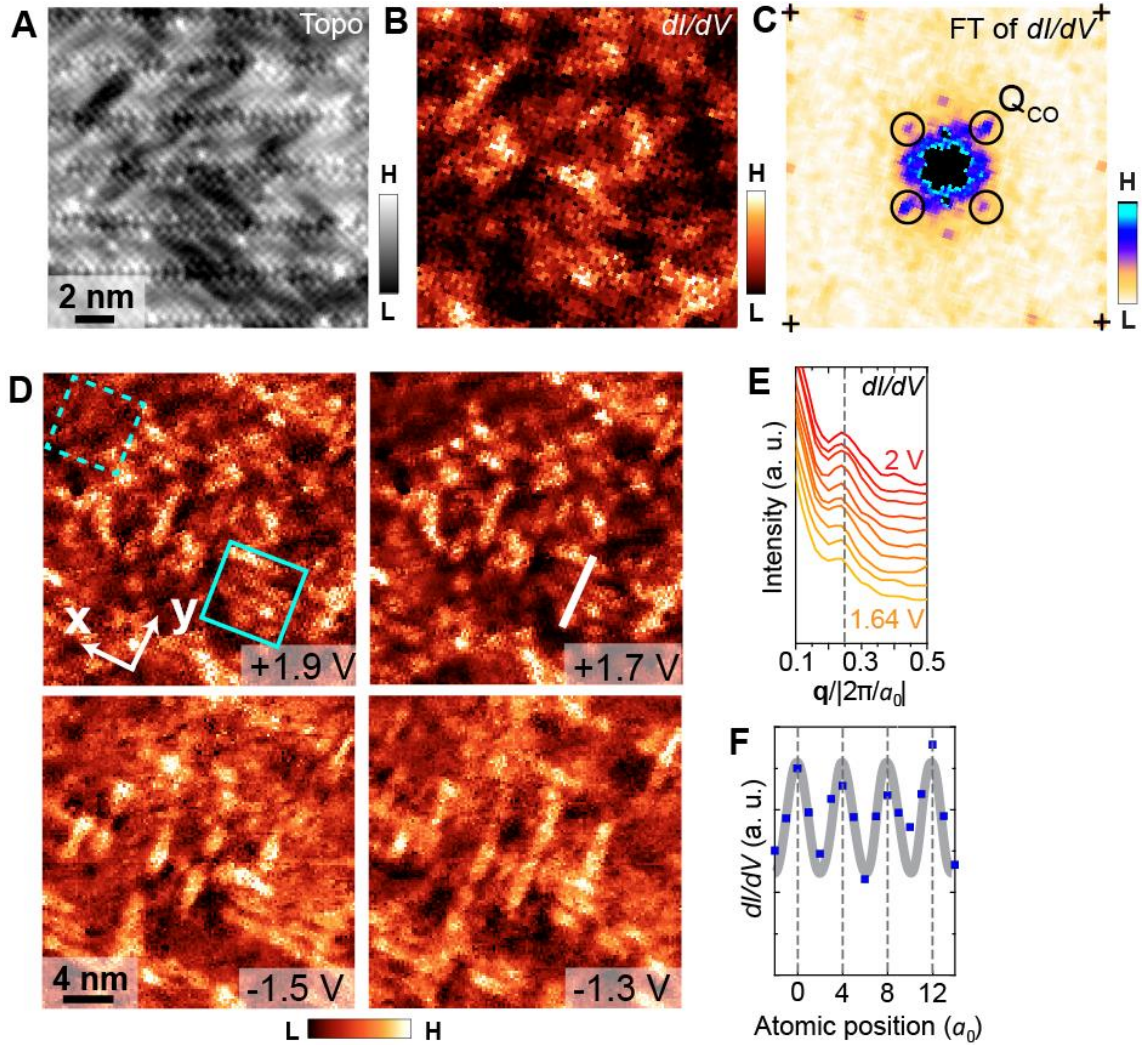

**Figure 3. Visualizing the charge stripe order in the STM conductance maps of insulating Bi-2212.** (A) Atomically-resolved STM topograph and (B) $dI/dV(\mathbf{r}, V = 2\text{ V})$ map acquired over the same ~15 nm square region of the sample. (C) Discrete FT of (B) showing $4a_0$ charge modulation peaks along $\mathbf{Q}_x$ and $\mathbf{Q}_y$. Small crosses in each corner denote the positions of the atomic Bragg peaks. (D) Four integrated $dI/dV(\mathbf{r}, V)$ maps at different STM biases acquired over a ~25 nm square region of the sample. To enhance signal-to-noise, each map is an average of 5 $dI/dV$ conductance maps, spaced by 40 mV, centered on the bias labeled. (E) A series of linecuts, offset for clarity, in the 4-fold symmetrized FTs of $dI/dV(\mathbf{r}, V)$ maps at 10 equally spaced biases from V = 1.64 V to 2 V. (F) A real-space linecut along the white line in (D) that shows the periodic variation of $dI/dV$ conductance in the $dI/dV$ map (blue squares), and the ideal $4a_0$ modulation as a visual guide (gray line). STM setup: $V_{sample}$= - 2 V, $I_{set}$ = 20 pA and $V_{exc}$=20 mV.

## Methods

The starting point for our experiment is a single crystal of optimally-doped Bi-2212 ($T_c \sim 91$ K), grown using travelling floating zone method. This crystal is cleaved in UHV at room-temperature between adjacent BiO layers, and imaged by STM at ~4.5 K to demonstrate that the STM topograph and *dI/dV* spectra are consistent with previous work (for example, Fig. 1B of the main text). Then, the same sample (without re-cleaving) is taken out of the STM, inserted into the annealing stage in the adjacent UHV chamber, and heated up (annealed) at varying temperatures to achieve the doping levels presented in the paper. Annealing is performed by the standard procedure for substrate heating also used during molecular beam epitaxy growth. More specifically, heating is achieved by running a current through a heating filament (made from a W wire) inside the sample holder, positioned ~2-3 mm behind the sample plate. The temperature of the sample plate is calibrated before, as well as after the completion of the STM portion of the experiment in an external UHV chamber, by attaching a thermocouple directly to the top of the sample plate. The sample reaches ~85% of the final temperature within the first ~5 minutes of heating, and the final temperature that the sample reaches is the value reported in the main text. For example, the sample in Fig. 1C was annealed for 10 minutes to reach the maximum temperature of ~270 °C, and that in Fig. 1D was annealed for 3 cycles of 20 minutes each, reaching the maximum of ~380 °C (a single cycle of 20 minute heating gave qualitatively the same results). We emphasize again that the sample is <u>never taken out of the UHV environment</u> from the time it is first cleaved, until the STM experiments are completed. Annealing at temperatures of ~430 °C (4 attempts) yielded an insulating surface that we could not tunnel into at ~4.5 K using STM setup voltage as high as 3 V and the tunneling current as low as ~ few pA, despite no change in the room-temperature RHEED image of the surface (Figs. S1, S2). This is consistent with more oxygens escaping from the several topmost layers, which further underdopes the surface, without affecting the structural morphology of the surface.

We note that the bulk of our UHV annealed Bi-2212 samples still exhibit superconductivity, with $T_c$ reduced by only ~2-3% percent (Figure S1), in spite of the dramatic change of the electronic properties at the surface (Fig. 1B-D). Thus, the effects observed here are a consequence of surface doping, possibly of only several topmost layers where oxygen dopant concentration is significantly reduced. Nevertheless, given the surface-sensitivity of STM, the measurements acquired on this insulating surface are not expected to be affected by the conducting bulk, as further discussed in Supplementary Information 3.

Although difficult to quantify, the rate of oxygen dopants escaping during annealing is likely not a linear function of temperature or the annealing time, as the superconducting bulk acts as an infinite reservoir of oxygen dopants that can partially replace some of the interstitials that escape from the surface. Nevertheless, based on our data of dopant densities and electronic properties of annealed samples, we can conclude that the surface doping obtained in the end follows the maximum temperature of the annealing used.

All STM data was acquired at the base temperature of ~4.5 K. Spectroscopic measurements have been taken using a standard lock-in technique at 915 Hz frequency and varying bias excitation as detailed in the figure captions. The STM tips used were home-made, chemically etched W tips annealed to bright-orange color in UHV. Tip quality has been evaluated on the surface of a single crystal Cu(111) prior to performing the measurements presented in this paper. The Cu(111) surface was cleaned by repeated cycles of heating and Argon sputtering in UHV before it was inserted into the STM head.

**Code availability**

The computer code used for data analysis is available upon request from the corresponding author.

**Data availability**

The data supporting the findings of this study are available upon request from the corresponding author.

**Acknowledgements:**

We thank M. Allan, J. C. Davis, J. E. Hoffman, S. Kivelson and S. Sachdev for insightful conversations. I.Z. gratefully acknowledges the support from DARPA Grant Number N66001-17-1-4051, Army Research Office Grant Number W911NF-17-1-0399, National Science Foundation Grant Number NSF-DMR-1654041 and the Boston College startup for the partial support of H.Z, Z.R. and B.R. Z.W. acknowledges the support from U.S. Department of Energy, Basic Energy Sciences Grant No. DE-FG02-99ER45747. The work in Brookhaven was supported by the Office of Science, U.S. Department of Energy under Contract No. DE-SC0012704. J.S. and R.D.Z. were supported by the Center for Emergent Superconductivity, an Energy Frontier Research Center funded by the U.S. Department of Energy, Office of Science.